\documentclass[conference]{IEEEtran}
\IEEEoverridecommandlockouts
\usepackage{cite}
\usepackage{amsmath,amssymb,amsfonts}
\usepackage{algorithmic}
\usepackage{graphicx}
\usepackage[caption=false,font=footnotesize]{subfig}
\usepackage{textcomp}
\usepackage{xcolor}
\usepackage{url}
\usepackage{booktabs}
\usepackage{multirow}
\def\BibTeX{{\rm B\kern-.05em{\sc i\kern-.025em b}\kern-.08em
    T\kern-.1667em\lower.7ex\hbox{E}\kern-.125emX}}
\begin{document}

\title{HiMark: Hierarchical and Controllable Multi-bit Watermarking for Large Language Models}

\author{
\IEEEauthorblockN{Ke Zhu$^1$, Yanli Chen$^{1,\star}$, Ling Xiong$^2$, Gejian Zhao$^3$ and Hanzhou Wu$^{1,3,\star}$}
\IEEEauthorblockA{
$^1$School of Big Data and Computer Science, Guizhou Normal University, Guiyang 550025, China\\
$^2$School of Information Science, Beijing Language and Culture University, Beijing 100083, China\\
$^3$School of Communication and Information Engineering,
Shanghai University, Shanghai 200444, China\\
$^\star$Corresponding authors (e-mails: yanli\_027@163.com, h.wu.phd@ieee.org)}}

\maketitle

\begin{abstract}
With the rapid adoption of large language models (LLMs) in text generation applications, it has become increasingly difficult to trace generated content and assign responsibility for its use, raising concerns about provenance, copyright protection, and accountability. Multi-bit watermarking addresses these issues by embedding multi-bit provenance information into generated text, but reliable extraction remains difficult due to sparse per-bit evidence and the trade-off between watermark-extraction accuracy and text quality. To deal with these challenges, we propose HiMark, a hierarchical and controllable multi-bit watermarking framework for LLM-generated text. During watermark embedding, HiMark uses a hierarchical candidate-control mechanism, including initial candidate sampling, layer-wise tournament elimination, support-based refinement, and support-threshold filtering. During watermark detection, a model-agnostic evidence aggregation mechanism computes multi-layer support evidence for each observed token, and then accumulates these token-level votes into the corresponding information for robust watermark extraction. Experiments show that HiMark improves watermark extraction accuracy while maintaining text quality, e.g., the watermark extraction accuracy is 90.01\% under the paraphrasing attack when 32 bits were embedded within a text containing 300 tokens, demonstrating its effectiveness for provenance tracking.
\end{abstract}


\section{Introduction}
Large language models (LLMs) have already made remarkable progress and demonstrated strong capabilities in question answering, text writing, code generation, and intelligent assistant systems. The widespread deployment of LLMs, however, has also introduced a series of risks such as copyright infringement, misinformation dissemination, malicious content generation, and synthetic data contamination. Therefore, verifiable provenance marking is essential for detecting LLM-generated content and source tracing in practical LLM applications.

Model watermarking~\cite{Yang2025Mat} has received extensive attention as a feasible solution. It embeds a secret watermark into the generated text for provenance verification. Mainstream methods focus on zero-bit verification, where the watermark is used to determine the presence of a hidden pattern in a given text, rather than to extract a long message out. Although zero-bit watermarking is effective for distinguishing watermarked text from unwatermarked one, it could not extract additional provenance information such as user identities and model IDs.

To tackle this limitation, recent studies have begun to focus on multi-bit watermarking~\cite{Yoo2023AdvancingBI,Wang2023TowardsCW}. Compared with zero-bit watermarking, multi-bit watermarking enables us to extract more information from the watermarked text. A very common strategy to realize multi-bit watermarking is using different token positions to carry different message bits, which allows each bit to be extracted by collecting evidence from a limited portion of generated tokens. As the number of to-be-embedded bits increases, per-bit evidence becomes fragile, especially making extraction less stable in short texts. Although heavy interventions on the base language model enhance watermark extraction precision, they degrade the quality of the generated text~\cite{Kirchenbauer2024Reliability}. While weaker interventions maintain fluency but may weaken the performance of watermark extraction. 

We introduce HiMark, a hierarchical and controllable multi-bit watermarking framework for LLM-generated text. HiMark does not require LLM training, fine-tuning, and other internal interventions. Instead, it guides output token selection within the local candidate token set produced by the LLM through a candidate-control mechanism, enabling robust multi-bit embedding and extraction. Extensive experimental results confirm the superiority. The main contributions of this paper include:

\begin{itemize}
    \item We introduce a novel watermarking framework for LLM-generated text. The proposed framework enables multi-bit watermark embedding and extraction without altering the LLM, and possesses strong robustness in short-text, long-message, and textual perturbation scenarios.
    
    \item We design a method shifting watermark embedding from global probability intervention to local token selection. It samples local candidates and guides them through layer-wise tournament elimination, support-based refinement, and support-threshold filtering, allowing the most suitable tokens to be selected to carry the corresponding bits. 

    \item In order to achieve model-agnostic watermark extraction, we propose a two-stage evidence aggregation strategy that evaluates the multi-layer support of each token for bit `0' or `1' and accumulates evidence across multiple tokens into a voting matrix, thereby enhancing the robustness of multi-bit extraction due to textual perturbations.

    \item Experimental results show that the proposed method not only ensures that the generated texts are with high quality, but also enables the embedded multi-bit information to be reliably extracted for intellectual property protection.
\end{itemize}

\section{Related Works}
LLM watermarking \cite{Yang2025Mat} can be divided into post-hoc watermarking and inference-time watermarking. The former embeds information by modifying the generated texts, through lexical substitution, synonym replacement, or syntactic rewriting~\cite{Yang2021TracingTP,Yoo2023RobustMN,Ding2025LOCATLT}, which require additional editing, thereby affecting semantic consistency or text quality. The embedded information can be weakened by subsequent editing or paraphrasing attacks.

We focus on inference-time watermarking~\cite{Kirchenbauer2023AWF,Kuditipudi2023RobustDW,Dathathri2024ScalableWF}, which embeds watermarks during text generation by intervening in token sampling or decoding. Both post-hoc watermarking and inference-time watermarking can be further categorized into zero-bit watermarking and multi-bit watermarking. 

\subsection{Zero-bit Watermarking}
Zero-bit watermarking~\cite{Kirchenbauer2024Reliability} determines whether a specified pattern is present in a text. Existing techniques can be broadly divided into probability-biased watermarking, unbiased watermarking, and sampling-based watermarking. Among them, the probability biasing~\cite{Kirchenbauer2023AWF} is the most popular, i.e., partitioning the vocabulary into green and red sets, increasing the probability of selecting green-list tokens during generation, and finally extracting the watermark based on green-list token frequency. However, such interventions in the base language model may affect text quality. For example, simply increasing the number of green-list tokens or the probability of green-list tokens as output may increase the probability of rare tokens as output, reducing the naturalness of generated text.

To reduce the impact, unbiased watermarking methods are proposed so as to keep the watermarking intervention unbiased mathematically, e.g., Hu \emph{et al}.~\cite{Hu2023UnbiasedWF} combined inverse sampling with $\gamma$-reweighting for watermark embedding. The approach in \cite{wu2024resilientaccessibledistributionpreservingwatermark} further improved $\gamma$-reweighting. Nevertheless, enforcing unbiasedness in expectation may reduce extraction accuracy as it restricts the strength of watermark intervention and makes the generated tokens closer to the original sampling behavior, leaving weaker statistical evidence for detection. 

Sampling-based watermarking provides another solution by embedding watermarks through token sampling. For example, Kuditipudi \emph{et al}.~\cite{Kuditipudi2023RobustDW} proposed a distortion-free watermarking scheme that uses a secret key to control sampling randomness, and meanwhile SynthID-Text~\cite{Dathathri2024ScalableWF} introduced tournament sampling to select tokens through layer-wise randomized comparisons. The same randomized rules are applied during detection, and token-level scores are accumulated to determine whether the text is watermarked. HiMark is inspired by the tournament-style token selection mechanism of SynthID-Text, but differs in both objective and design. SynthID-Text targets zero-bit watermark detection, while HiMark extends tournament-based token selection to multi-bit watermarking by introducing multilayer bit-specific evidence, message-index assignment, support-based refinement, and threshold filtering.

\subsection{Multi-bit Watermarking}
Common multi-bit watermarking methods include message coupled, segment based, and position allocation. Specifically, message coupled methods~\cite{Wang2023TowardsCW} embed the message by incorporating it into the generated watermarking rule. The watermark detector typically needs to match, search, or enumerate candidate messages to identify the most likely embedded message. Although this strategy embeds full information, its extraction cost grows exponentially with the message length, accordingly limiting efficient long-message extraction.

Segment based techniques~\cite{lin2025dermarkdynamicefficientrobust} partition a text into multiple disjoint segments, each generally carrying one bit. However, segment-level extraction still depends on sufficiently long and well-aligned text segments. When the text is short or large portions are removed, some segments may become difficult to align, leading to unstable message extraction.

Position allocation methods assign different message bits to different token positions and extract the message by aggregating token-level evidence. E.g., MPAC~\cite{Yoo2023AdvancingBI} uses green-list-based probability biasing, while BiMark~\cite{Feng2025BiMarkUM} introduces unbiased multi-layer reweighting to ensure unbiasedness. However, for multi-bit extraction, each bit can only collect evidence from a limited subset of generated tokens. As the message length increases or the text becomes shorter, the evidence for each bit becomes sparse, making robust extraction more challenging. It therefore motivates us to find novel watermarking strategies.

\begin{figure*}[!ht]
    \centering
    \includegraphics[
        width=0.8\linewidth,
        height=0.45\linewidth
    ]{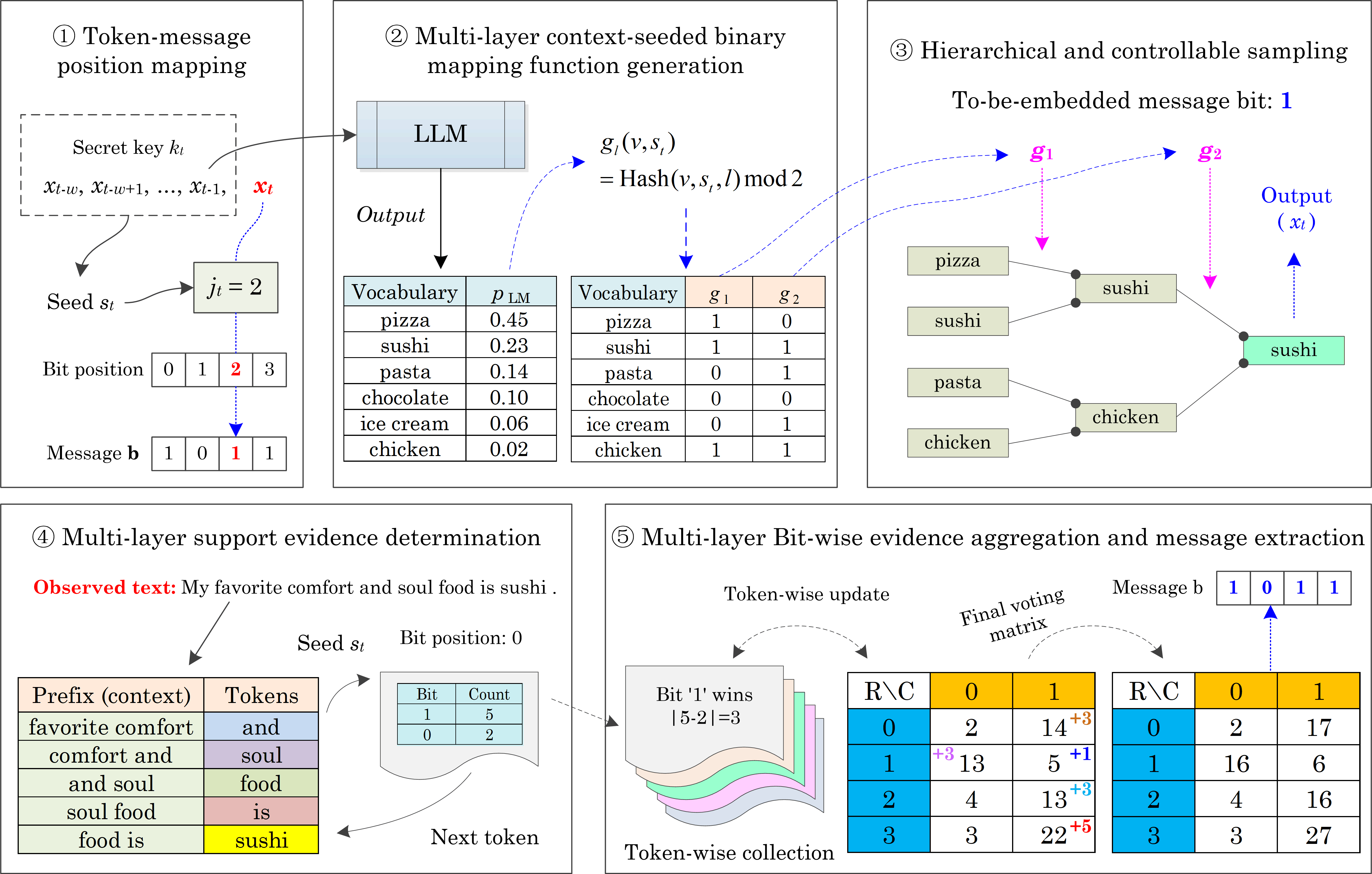}
    \caption{Overview of the proposed framework: The first three steps correspond to watermark embedding, while the last two correspond to watermark extraction.}
    \label{fig:watermark_framework}
\end{figure*}
\section{Proposed Method}
HiMark enhances the watermarking evidence through hierarchical candidate selection during embedding and two-stage evidence aggregation during extraction. During watermark embedding, given a secret message, a secret key, and the current prefix context, HiMark determines the message bit assigned to the current token, and uses hierarchical and controllable sampling to select a candidate token that provides strong support for the target bit. During watermark extraction, the detector reconstructs the message-position mapping. Through a two-stage evidence aggregation strategy, the resulting evidence is accumulated into the message-bit positions. Finally, the binary message is extracted. The overall framework refers to Fig.~\ref{fig:watermark_framework}.

\begin{figure}[!ht]
    \centering
    \includegraphics[width=\linewidth]{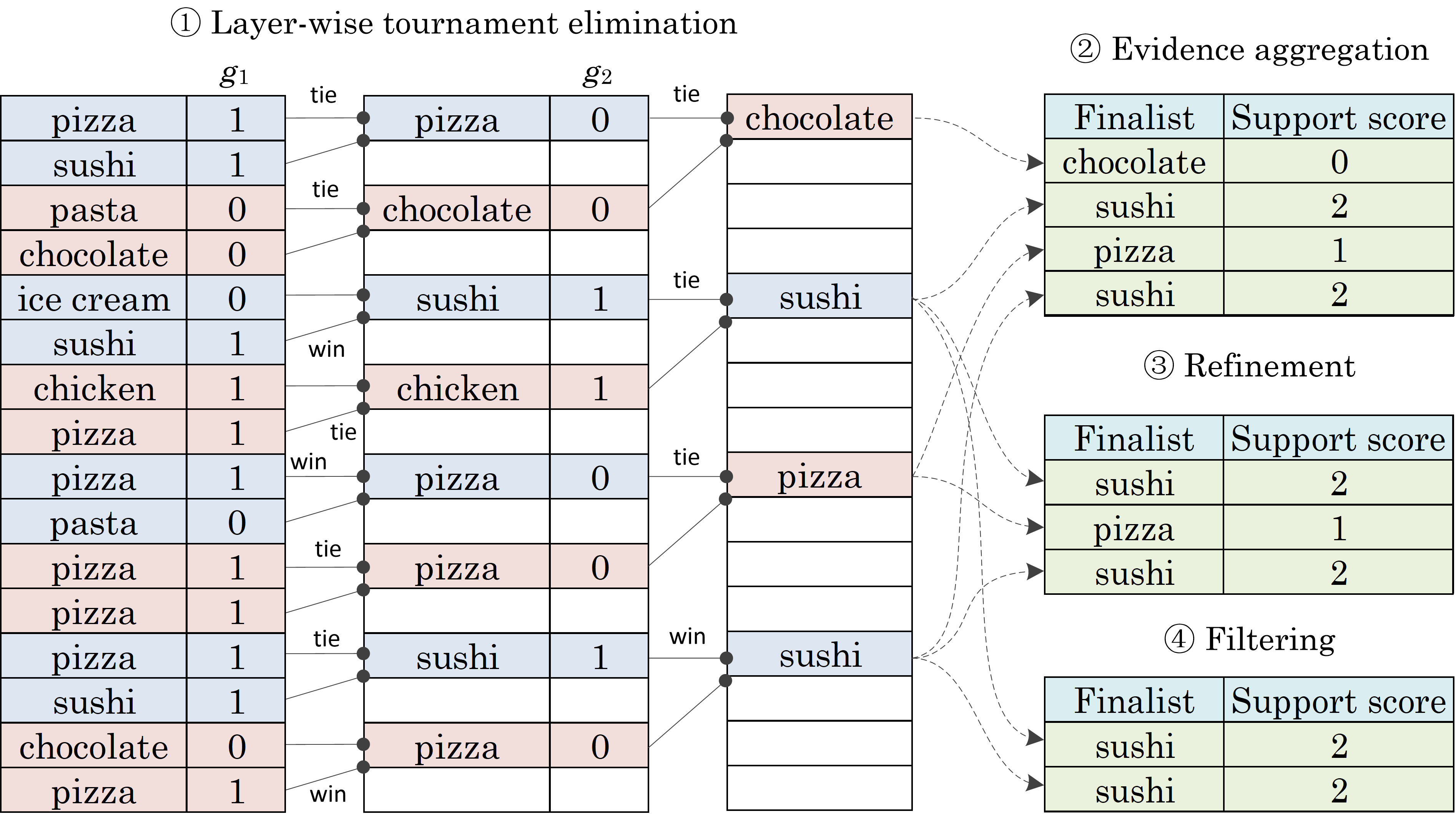}
    \caption{Illustration for hierarchical and controllable sampling.}
    \label{fig:sampling}
\end{figure}

\subsection{Hierarchical and Controllable Watermark Embedding}
Given a binary message, the watermark embedding process selects an output token that carries the assigned message bit while maintaining the generation preference of the LLM. At each generation step, HiMark first uses the prefix context and secret key to determine which message-bit position should be embedded. It then samples candidate tokens from the original LLM distribution and evaluates whether each candidate supports the selected bit. Finally, HiMark progressively filters the candidates through layer-wise tournament elimination, support-based refinement, and support-threshold filtering, and samples the output token from the refined candidate set.

Mathematically, let $\textbf{b}\in \{0,1\}^L$ denote the secret message, where $b_i\in\{0,1\}$ is the $i$-th bit and $L$ is the length. At token generation step $t > 0$, HiMark collects the prefix context by $p_t = (x_{t-w}, x_{t-w+1}, ..., x_{t-1})$, where $x_t$ denotes the token to be generated at step $t$ and $w$ is the context window-size. The prefix context $p_t$ together with a secret key $k_t$ will be mapped to a context-dependent seed, expressed as $s_t = \text{PRF}(p_t, k_t)$, where $\text{PRF}$ is a key-controlled deterministic function. $s_t$ allows us to generate an index $j_t = f(s_t)\in [0, L)$ that specifies the bit to be embedded at step $t$. The corresponding bit, i.e., $b_{j_t}$, is defined as the target bit $y_t$. It is noted that the above operation affects the evidence coverage of embedded message. In detail, let $n$ be the number of effective tokens generated. For any bit $b_i$, the probability that it does not receive a particular evidence is $1-1/L$. The probability that no evidence is given to $b_i$ is $(1-1/L)^n$, so that the total probabilities of all bits receiving no evidence is $L(1-1/L)^n \leq Le^{-n/L}$. Therefore, covering all bits with probability at least $1-\delta$ is guaranteed when
\begin{equation}
n\geq L\cdot\text{ln}(L/\delta),
\end{equation}
which implies that longer messages require more effective tokens and longer texts improve watermark-extraction reliability.

Let $p_\text{LM}(\cdot|x_{<t})$ denote the next-token distribution produced by the LLM at generation step $t$. HiMark samples an initial candidate set $C_t^{(0)}$ containing $2^m$ tokens from this distribution, where $m$ denotes the tournament depth. To evaluate whether a candidate token supports the target bit $y_t$, HiMark defines a context-seeded binary mapping function as follows:
\begin{equation}
g_l(v, s_t) = \text{Hash}(v, s_t, l)~\text{mod}~2,
\end{equation}
where $v$ represents the ID of candidate token, and $l\in [1,m]$ denotes the layer index. The candidate token $v$ supports the target bit $y_t$ at layer $l$ if and only if $g_l(v,s_t)=y_t$.

We now perform hierarchical candidate selection, including layer-wise tournament elimination, support-based refinement, and support-threshold filtering. As illustrated in Fig.~\ref{fig:sampling}, during layer-wise tournament elimination, candidate tokens are paired according to their sampled order, and the candidate supporting the target bit advances to the next layer. For a candidate pair $(u,v)$, the winner at layer $l\in [1,m]$ is determined as
\begin{equation}
\text{win}_l(u,v)=
\begin{cases}
u,\text{if}~g_l(u,s_t)=y_t,g_l(v,s_t)\neq y_t,\\
v,\text{if}~g_l(u,s_t)\neq y_t,g_l(v,s_t)=y_t,\\
\text{Random}(u,v),\text{if}~g_l(u,s_t)=g_l(v,s_t).
\end{cases}
\end{equation}

The tournament elimination stops when the size of candidate set is reduced to $K_\text{keep}$. For each retained candidate token $v$, HiMark computes its support score as follows:
\begin{equation}
\text{support}(v) = \sum_{l=1}^{m}\mathbb{I}[g_l(v,s_t) = y_t],
\end{equation}
where $\mathbb{I}[\cdot]$ means the indicator function. HiMark retains the top $K_\text{refine}$ candidates with the highest support scores and removes candidates whose support scores are lower than $\tau_s$.  Discarded candidates are all masked by setting their logits to $-\infty$, while the remaining candidates keep the original LLM logits. The output token is sampled from the refined candidate set.

\subsection{Two-Stage Evidence Aggregation for Watermark Extraction}
During watermark extraction, we extract the binary message from the observed text without accessing the original LLM. This process uses a two-stage evidence aggregation strategy, consisting of multi-layer token support computation and bit-wise evidence aggregation. The former reconstructs the message index of each observed token and computes its support for the corresponding bit. The latter accumulates token-level evidence into a voting matrix and extracts the binary message.

In detail, for each token $x_t$, the detector constructs the prefix context $p_t$ from the observed text and uses the secret key $k_t$ to reconstruct the context-dependent seed $s_t$ and message index $j_t$ by $s_t = \text{PRF}(p_t, k_t)$ and $j_t = f(s_t)\in [0, L)$. Then, the detector evaluates $x_t$ using the identical context-seeded binary function as that used during watermark embedding, i.e.,
\begin{equation}
g_l(x_t, s_t) = \text{Hash}(x_t, s_t, l)~\text{mod}~2,~\forall l\in [1,m].
\end{equation}
The support counts for the corresponding token are 
\begin{equation}
n_t^{(1)}=\sum_{l=1}^{m} g_l(x_t,s_t),~n_t^{(0)}=m-n_t^{(1)}.
\end{equation}
The larger support count determines the vote direction of $x_t$, and the absolute support difference gives the evidence strength.

For bit-wise evidence aggregation, HiMark initializes $\textbf{C} = \{0\}^{L\times2}$ , where each row corresponds to a message-bit position and the two columns store the accumulated evidence for the corresponding bit. Each token contributes evidence to the position indicated by its reconstructed index $j_t$. Specifically, the larger one between $n_t^{(0)}$ and $n_t^{(1)}$ determines the voted bit, and the absolute difference between them is accumulated as the evidence weight in the column of the voted bit. After all observed tokens are processed, the binary message is extracted through weighted voting at each message-bit position. For each message-bit position $j$, the detector selects the bit value with the larger accumulated evidence in the voting matrix:
\begin{equation}
b_j' = \arg\max_{d\in\{0,1\}}~\textbf{C}[j,d],~\forall j\in [0,L),
\end{equation}
by which we can reconstruct the full message $\textbf{b}' = \{b_j'\}_{j=0}^{L-1}$.

\section{Experimental Results and Analysis}
We use Llama-3.1-8B\footnote{Online available: \url{https://huggingface.co/meta-llama/Llama-3.1-8B}} as the base generation model, which is a decoder-only autoregressive LLM. The secret message is a random binary string. We sample prompts from C4 dataset\footnote{Online available: \url{https://huggingface.co/datasets/allenai/c4}}, which is a cleaned English web corpus derived from Common Crawl and available through HuggingFace Datasets.

The text quality is measured by perplexity (PPL). All PPLs are computed using Llama-3.1-8B and its tokenizer, with the prompt tokens excluded from the PPL calculation. Specifically, PPL is calculated on the generated token sequence according to the standard autoregressive language-model likelihood, where a lower PPL indicates more fluent of the generated text.

For multi-bit watermark extraction, we compare our method with MPAC~\cite{Yoo2023AdvancingBI} and BiMark~\cite{Feng2025BiMarkUM}. For MPAC, e.g., $\delta = 1.5$ means the watermark strength. For zero-bit watermark detection, we here include Soft Red List~\cite{Kirchenbauer2023AWF} and SynthID-Text~\cite{Dathathri2024ScalableWF}. Unless otherwise specified, HiMark embeds an 8-bit random string and uses $w=2$ and $m=11$, where $m$ is the tournament depth and corresponds to $2^{11}=2048$ initial candidates. Other parameters are $K_\text{keep}=32$, $K_\text{refine}=10$, and $\tau_s=5$. Unless otherwise specified, each experimental result is averaged over 1,000 generated samples. Moreover, all generation steps share the identical $k_t$.

\begin{table}[t]
\centering
\caption{Watermark extraction accuracy (\%) and perplexity (PPL) under different message lengths and text lengths.}
\label{tab:multi_bit_results}
\fontsize{7pt}{8pt}\selectfont
\setlength{\tabcolsep}{4.0pt}
\renewcommand{\arraystretch}{1.0}
\begin{tabular}{c|c|cccccccc}
\hline
\multirow{3}{*}{Bits} & \multirow{3}{*}{Method}
& \multicolumn{8}{c}{Text length (tokens)} \\
\cline{3-10}
& & \multicolumn{2}{c}{50}
& \multicolumn{2}{c}{100}
& \multicolumn{2}{c}{200}
& \multicolumn{2}{c}{300} \\
\cline{3-10}
& & Acc. & PPL & Acc. & PPL & Acc. & PPL & Acc. & PPL \\
\hline

\multirow{4}{*}{8}
& MPAC (1)   & 49.49 & 9.44 & 79.72 & 8.64 & 89.51 & 8.08 & 93.48 & 7.85 \\
& MPAC (1.5) & 78.81 & 9.76 & 89.75 & 9.13 & 96.40 & 8.63 & 98.57 & 8.39 \\
& BiMark     & 95.26 & 8.50 & 97.62 & 7.38 & 98.15 & 5.78 & 97.88 & \textbf{4.57} \\
& HiMark     & \textbf{97.36} & \textbf{6.25}
             & \textbf{99.18} & \textbf{5.83}
             & \textbf{99.64} & \textbf{5.39}
             & \textbf{99.71} & 5.11 \\
\hline

\multirow{4}{*}{16}
& MPAC (1)   & 57.04 & 9.36 & 68.25 & 8.63 & 79.31 & 8.17 & 87.50 & 7.89 \\
& MPAC (1.5) & 66.06 & 9.69 & 78.21 & 8.87 & 89.04 & 8.45 & 93.83 & 8.45 \\
& BiMark     & 85.55 & 8.60 & 93.31 & 7.35 & 95.54 & 5.83 & 95.54 & \textbf{4.71} \\
& HiMark     & \textbf{91.70} & \textbf{6.18}
             & \textbf{96.23} & \textbf{5.74}
             & \textbf{98.86} & \textbf{5.32}
             & \textbf{99.28} & 5.02 \\
\hline

\multirow{4}{*}{32}
& MPAC (1)   & 45.06 & 9.16 & 56.79 & 8.34 & 67.96 & 7.92 & 74.32 & 7.67 \\
& MPAC (1.5) & 51.03 & 9.77 & 65.35 & 8.96 & 78.15 & 8.49 & 84.70 & 8.35 \\
& BiMark     & 66.35 & 8.63 & 82.69 & 7.48 & 89.68 & 5.89 & 90.22 & \textbf{4.75} \\
& HiMark     & \textbf{81.93} & \textbf{6.68}
             & \textbf{90.18} & \textbf{6.13}
             & \textbf{95.57} & \textbf{5.57}
             & \textbf{97.32} & 5.28 \\
\hline
\end{tabular}
\vspace{0.5mm}
\noindent
\end{table}

\begin{table}[t]
\centering
\caption{Examples generated by different watermarking methods.}
\label{tab:ppl_quality_examples}
\scriptsize
\setlength{\tabcolsep}{3pt}
\renewcommand{\arraystretch}{1.0}
\begin{tabular}{c|p{0.76\linewidth}}
\hline
Method & Example excerpt \\
\hline
MPAC (1)
& ``The consolidated report is to be sent to the Ministry headquarters on April 23, the report is to be sent to the Ministry headquarters...'' \\
\hline
MPAC (1.5)
& ``The man who served our country with grace, dignity and was, and remains, a man full of class. He was a man full of class...'' \\
\hline
BiMark
& ``He started the final two games of last season while the Browns were using Campbell...'' \\
\hline
HiMark
& ``An attack of diarrhea can cause dehydration, so it is important to eat foods that help to replace the fluids...'' \\
\hline
\end{tabular}
\vspace{-1mm}
\end{table}

\begin{table}[t]
\centering
\caption{The 8-bit watermark extraction accuracy (\%) across different datasets and LLM backbones.}
\label{tab:generalization_results}
\fontsize{7pt}{8pt}\selectfont
\setlength{\tabcolsep}{5.0pt}
\renewcommand{\arraystretch}{1.08}
\begin{tabular}{l|l|cccc}
\hline
\multirow{3}{*}{Setting}
& \multirow{3}{*}{Dataset / LLM backbone}
& \multicolumn{4}{c}{Watermark extraction accuracy (\%)} \\
\cline{3-6}
& & \multicolumn{4}{c}{Text length (tokens)} \\
\cline{3-6}
& & 50 & 100 & 200 & 300 \\
\hline

\multirow{3}{*}{Dataset}
& C4 / Llama-3.1-8B
& 97.36 & 99.18 & 99.64 & 99.71 \\

& OpenWebText / Llama-3.1-8B
& 98.01 & 99.65 & 99.90 & 99.95 \\

& CNN/DailyMail / Llama-3.1-8B
& 98.22 & 99.59 & 99.94 & 99.87 \\
\hline

\multirow{3}{*}{LLM}
& C4 / Llama-3.1-8B
& 97.36 & 99.18 & 99.64 & 99.71 \\

& C4 / Qwen2.5-7B-Instruct
& 95.95 & 98.17 & 99.42 & 99.71 \\

& C4 / OPT-2.7B
& 97.44 & 98.88 & 99.51 & 99.49 \\
\hline

\end{tabular}
\vspace{0.5mm}
\end{table}

\begin{table}[t]
\centering
\caption{Computational overhead comparison.}
\label{tab:computational_overhead}
\scriptsize
\setlength{\tabcolsep}{10pt}
\renewcommand{\arraystretch}{1.0}
\begin{tabular}{c|c|c}
\hline
Method & Avg. time per sample (s) & GPU memory (GB) \\
\hline
MPAC (1)   & 11.32 & \textbf{15.63} \\
MPAC (1.5) & 11.24 & 15.66 \\
BiMark     & \textbf{0.75} & 19.91 \\
HiMark     & 2.43 & 17.23 \\
\hline
\end{tabular}
\vspace{0.5mm}
\end{table}

\subsection{Watermark Extraction Accuracy and Text Quality}
As shown in Table~\ref{tab:multi_bit_results}, HiMark achieves the highest extraction accuracy across all settings by selecting tokens with stronger bit-level evidence for aggregation. It also maintains lower PPL than MPAC configurations and overall text quality superior to BiMark. Table~\ref{tab:ppl_quality_examples} presents representative 300-token examples of MPAC (1), MPAC (1.5), BiMark, and HiMark, with PPLs of 7.49, 8.34, 4.71, and 5.17, respectively. Higher-PPL texts show more repetition and awkward phrasing,  while HiMark remains more fluent, demonstrating  stronger watermark evidence with limited impact on fluency.

\begin{table}[t]
\centering
\caption{Watermark Extraction Accuracy (\%) under Synonym Substitution Attacks for an 8-Bit Random Binary String.}
\label{tab:synonym_attack}
\fontsize{7pt}{8pt}\selectfont
\setlength{\tabcolsep}{3.0pt}
\renewcommand{\arraystretch}{1.08}
\begin{tabular}{c|l|ccccccc}
\hline
\multirow{3}{*}{Ratio}
& \multirow{3}{*}{Method}
& \multicolumn{7}{c}{Watermark extraction accuracy (\%)} \\
\cline{3-9}
& & \multicolumn{7}{c}{Text length (tokens)} \\
\cline{3-9}
& & 25 & 50 & 100 & 150 & 200 & 250 & 300 \\
\hline

\multirow{4}{*}{0.1}
& MPAC (1)   & 50.38 & 59.50 & 67.08 & 73.41 & 76.34 & 80.15 & 82.10 \\
& MPAC (1.5) & 56.42 & 66.80 & 76.85 & 82.64 & 86.55 & 89.33 & 91.62 \\
& BiMark     & 70.25 & 82.40 & 91.02 & 93.74 & 94.94 & 95.23 & 95.24 \\
& HiMark     & \textbf{86.67} & \textbf{90.89} & \textbf{95.22}
& \textbf{97.07} & \textbf{97.75} & \textbf{98.32} & \textbf{98.66} \\
\hline

\multirow{4}{*}{0.2}
& MPAC (1)   & 45.51 & 52.83 & 60.72 & 64.29 & 67.31 & 70.03 & 72.19 \\
& MPAC (1.5) & 50.04 & 58.04 & 67.36 & 73.10 & 76.37 & 78.99 & 80.85 \\
& BiMark     & 60.61 & 71.60 & 81.71 & 86.29 & 88.62 & 89.39 & 90.01 \\
& HiMark     & \textbf{83.12} & \textbf{88.05} & \textbf{93.04}
& \textbf{95.40} & \textbf{96.84} & \textbf{97.48} & \textbf{98.07} \\
\hline

\multirow{4}{*}{0.3}
& MPAC (1)   & 43.18 & 48.71 & 53.36 & 56.30 & 58.29 & 59.78 & 60.57 \\
& MPAC (1.5) & 45.59 & 51.16 & 58.11 & 62.24 & 64.73 & 67.07 & 68.31 \\
& BiMark     & 53.37 & 62.52 & 70.37 & 74.22 & 76.23 & 77.30 & 78.23 \\
& HiMark     & \textbf{79.06} & \textbf{83.32} & \textbf{89.23}
& \textbf{92.59} & \textbf{94.08} & \textbf{95.17} & \textbf{96.23} \\
\hline
\end{tabular}
\end{table}

\begin{table}[t]
\centering
\caption{Watermark extraction accuracy (\%) under copy-paste attacks with 32 bits embedded within a text containing 400 tokens.}
\label{tab:copy_paste_attack}
\scriptsize
\setlength{\tabcolsep}{9pt}
\renewcommand{\arraystretch}{1.0}
\begin{tabular}{c|c|c|c}
\hline
\multirow{2}{*}{Method} & \multicolumn{3}{c}{Watermark extraction accuracy (\%)} \\
\cline{2-4}
& $\epsilon=0$ (no attack) & $\epsilon=0.2$ & $\epsilon=0.4$ \\
\hline
MPAC (1)   & 75.63 & 74.11 & 66.96 \\
MPAC (1.5) & 86.14 & 84.21 & 77.14 \\
BiMark     & 90.51 & 84.83 & 82.15 \\
HiMark     & \textbf{98.05} & \textbf{95.86} & \textbf{93.05} \\
\hline
\end{tabular}
\end{table}

\begin{table}[t]
\centering
\caption{The 8-bit watermark extraction accuracy (\%) under 30\% random deletion and insertion attacks.}
\label{tab:deletion_insertion_attack_results}
\fontsize{6.8pt}{7.6pt}\selectfont
\setlength{\tabcolsep}{6.4pt}
\renewcommand{\arraystretch}{1.0}
\begin{tabular}{l|l|ccccc}
\hline
\multirow{3}{*}{Attack} & \multirow{3}{*}{Method}
& \multicolumn{5}{c}{Watermark extraction accuracy (\%)} \\
\cline{3-7}
& & \multicolumn{5}{c}{Text length (tokens)} \\
\cline{3-7}
& & 25 & 50 & 100 & 200 & 300 \\
\hline

\multirow{4}{*}{Deletion}
& MPAC (1)   & 20.20 & 49.49 & 73.17 & 78.90 & 81.36 \\
& MPAC (1.5) & 67.82 & 75.00 & 80.92 & 87.96 & 89.89 \\
& BiMark     & 64.89 & 74.36 & 82.54 & 86.81 & 87.37 \\
& HiMark     & \textbf{90.56} & \textbf{91.44} & \textbf{94.06}
             & \textbf{94.19} & \textbf{94.03} \\
\hline

\multirow{4}{*}{Insertion}
& MPAC (1)   & 21.24 & 49.29 & 69.33 & 78.09 & 78.51 \\
& MPAC (1.5) & 69.15 & 75.58 & 86.84 & 89.81 & 93.75 \\
& BiMark     & 63.86 & 74.74 & 81.76 & 87.83 & 88.03 \\
& HiMark     & \textbf{84.75} & \textbf{86.69} & \textbf{91.25}
             & \textbf{94.06} & \textbf{94.69} \\
\hline
\end{tabular}
\vspace{0.5mm}
\end{table}

\begin{figure}[t]
    \centering
    \includegraphics[width=0.8\linewidth]{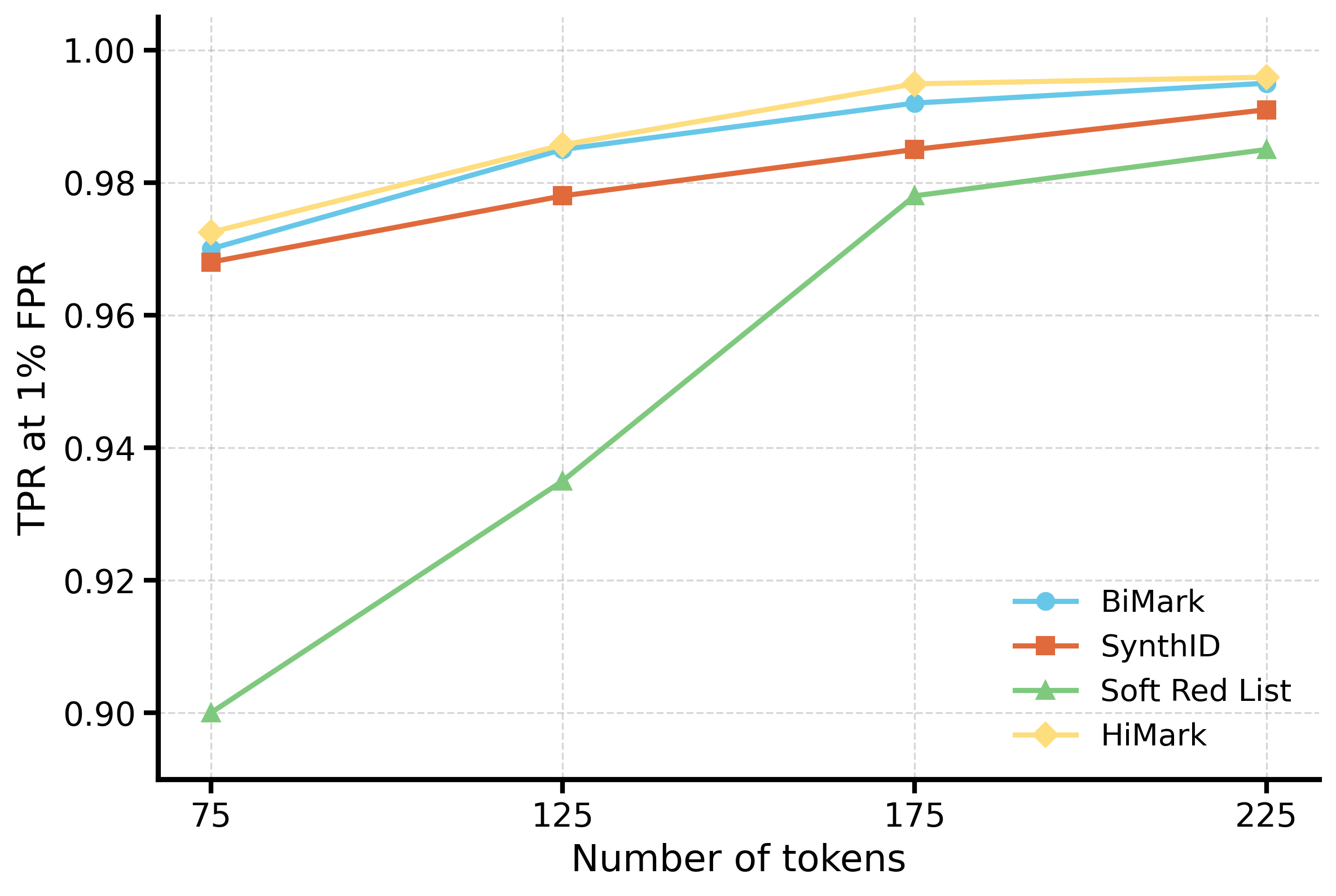}
    \caption{The zero-bit detection performance due to different text lengths.}
    \label{fig:zero_bit_detection}
    \vspace{-2mm}
\end{figure}

\begin{figure}[t]
    \centering

    \subfloat[Support threshold analysis]{
        \includegraphics[
            width=0.45\columnwidth,
            height=0.35\columnwidth,
            trim=5 5 5 5,
            clip
        ]{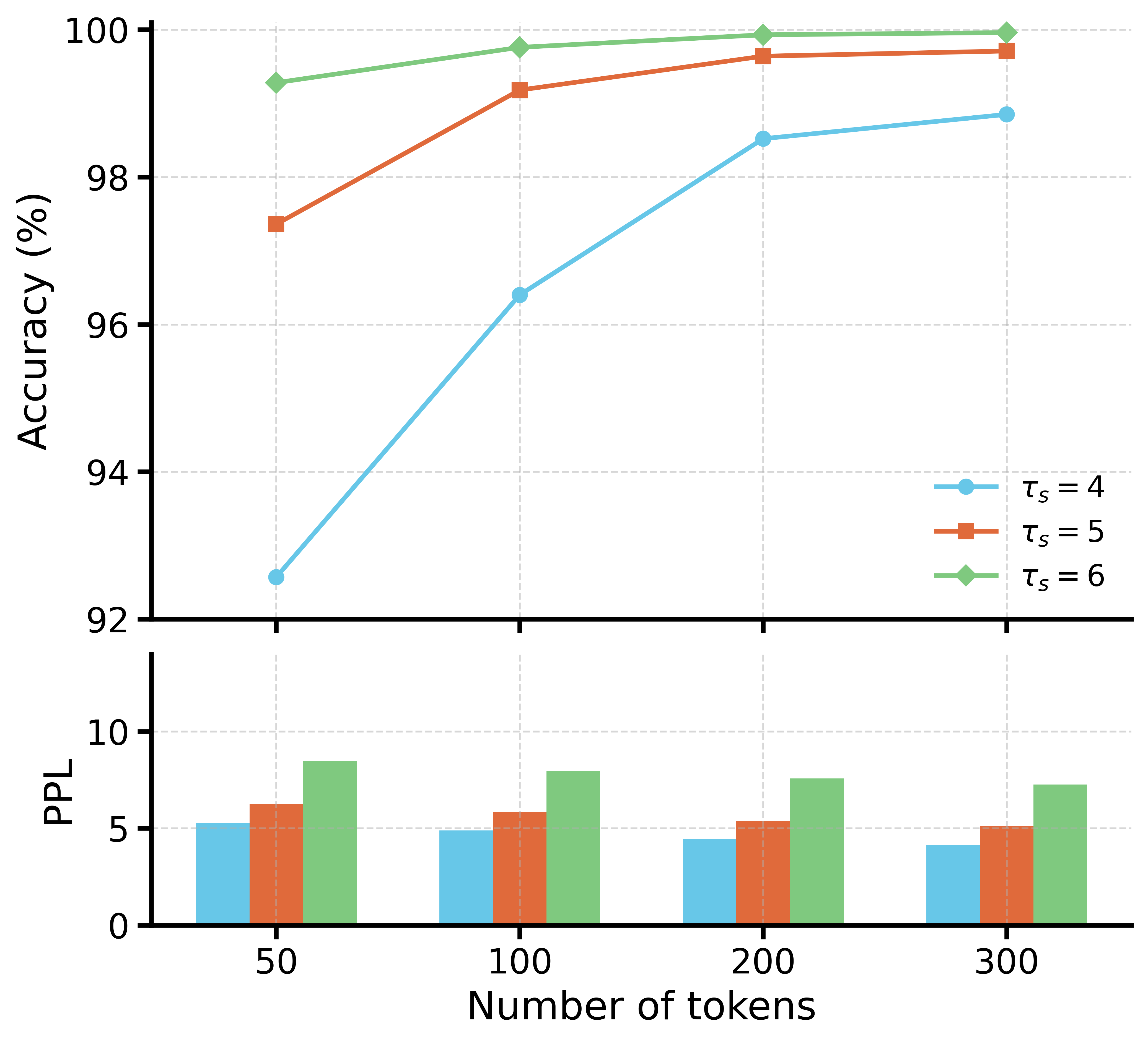}
        \label{fig:ablation_depth7_threshold}
    }
    \hfill
    \subfloat[Tournament depth analysis]{
        \includegraphics[
            width=0.45\columnwidth,
            height=0.35\columnwidth,
            trim=5 5 5 5,
            clip
        ]{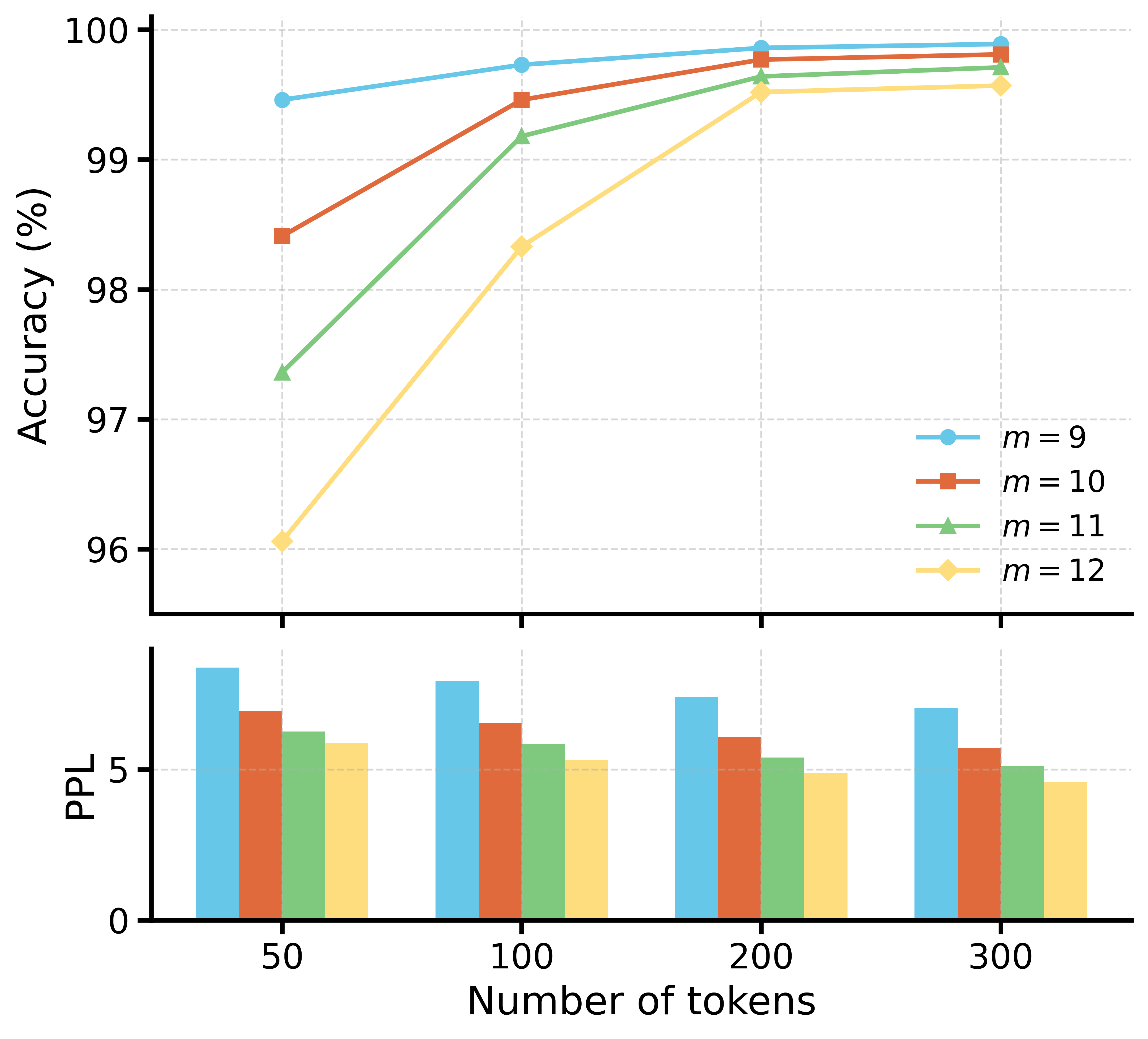}
        \label{fig:ablation_threshold2_depth}
    }

    \par\vspace{0.15cm}

    \subfloat[Refinement size analysis]{
        \includegraphics[
            width=0.45\columnwidth,
            height=0.35\columnwidth,
            trim=5 5 5 5,
            clip
        ]{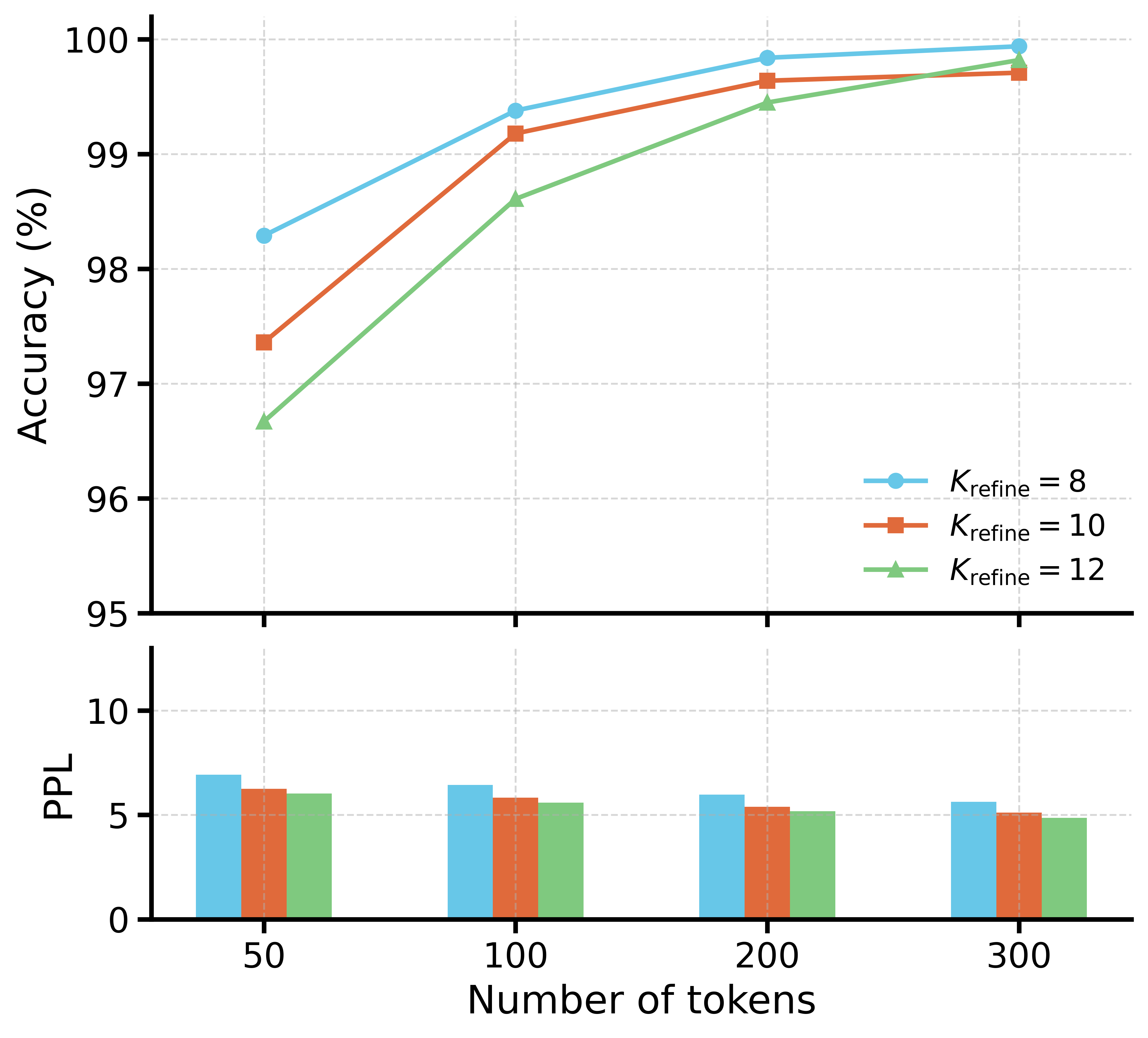}
        \label{fig:ablation_keep8_refine}
    }
    \hfill
    \subfloat[Candidate retention size analysis]{
        \includegraphics[
            width=0.45\columnwidth,
            height=0.35\columnwidth,
            trim=5 5 5 5,
            clip
        ]{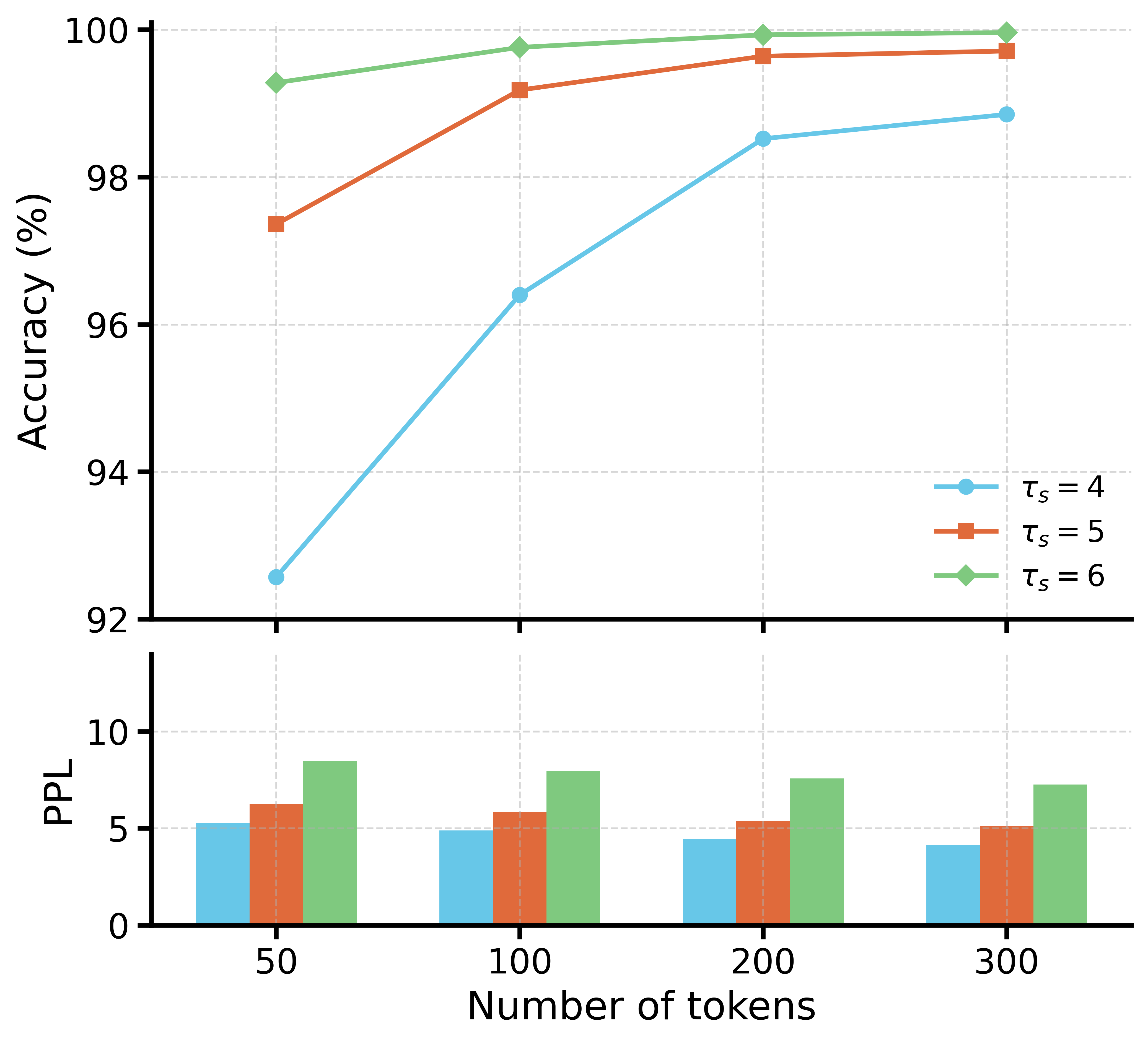}
        \label{fig:ablation_refine3_keep}
    }

    \caption{Ablation studies on the key hyperparameters of HiMark.
    (a) Effect of the support threshold $\tau_s$;
    (b) effect of the tournament depth $m$;
    (c) effect of the refinement size $K_{\text{refine}}$; and
    (d) effect of the candidate retention size $K_{\text{keep}}$.
    In each subfigure, the upper and lower panels report watermark extraction accuracy and PPL, respectively, under different generated text lengths.}
    \label{fig:ablation_all}
\end{figure}

\subsection{Generalization across LLMs and Datasets}
To verify generalizability, we evaluate HiMark across different LLMs and datasets. As shown in Table~\ref{tab:generalization_results}, HiMark achieves high extraction accuracy on Llama-3.1-8B, Qwen2.5-7B-Instruct\footnote{Online available: \url{https://huggingface.co/Qwen/Qwen2.5-7B-Instruct}}, OPT-2.7B\footnote{Online available: \url{https://huggingface.co/facebook/opt-2.7b}}, C4, CNN/DailyMail\footnote{Online available: \url{https://huggingface.co/datasets/abisee/cnn_dailymail}}, and OpenWebText\footnote{Online available: \url{https://huggingface.co/datasets/Skylion007/openwebtext}}, indicating that it is not tied to a specific model vocabulary size, or dataset distribution.

\subsection{Computational Overhead}
As shown in Table~\ref{tab:computational_overhead}, experiments evaluated on a single NVIDIA RTX 4090 GPU demonstrate that HiMark is slower than BiMark due to hierarchical candidate control but much faster than MPAC, with moderate GPU memory usage.

\subsection{Robustness of Multi-bit Extraction}
\subsubsection{Synonym Substitution Attack}
As shown in Table~\ref{tab:synonym_attack}, where the ratio denotes the proportion of randomly selected words replaced, HiMark remains robust under synonym substitution attacks generated using BERT~\cite{devlin2019bert} and WordNet~\cite{miller1995wordnet}. Although some tokens are replaced, many unchanged tokens still preserve valid watermark evidence, enabling stable extraction through multi-layer cross-token aggregation.

\subsubsection{Copy-Paste Attack}
As shown in Table~\ref{tab:copy_paste_attack}, where $\epsilon$ denotes the proportion of inserted unwatermarked content in the attacked text, HiMark remains robust under copy-paste attacks. Weighted evidence from unchanged watermarked regions enables reliable bit-wise voting despite watermark dilution.

\subsubsection{Robustness against Insertion and Deletion Attacks}
As shown in Table~\ref{tab:deletion_insertion_attack_results}, we randomly delete 30\% of tokens or insert 30\% additional tokens sampled from the tokenizer vocabulary at random positions. Despite disrupting prefix synchronization, HiMark maintains high extraction accuracy through evidence aggregation from unaffected regions.

\subsubsection{Spoofing/Forgery and Scrubbing Attacks}
We evaluate HiMark assuming attackers know the watermarking algorithm but not the legitimate key. For spoofing/forgery, attackers generate watermarked texts with independently chosen keys, while decoding uses the legitimate key, yielding 50.63\% bit accuracy. For scrubbing, attackers manually revise 300 watermarked texts to remove the watermark while preserving meaning, and HiMark retains 86.88\% extraction accuracy.

\subsubsection{Paraphrasing Attack}
For the paraphrasing attack, we use DIPPER paraphraser-xxl\footnote{Available: \url{https://huggingface.co/kalpeshk2011/dipper-paraphraser-xxl}} to rewrite 300-token watermarked texts containing 32-bit messages. We set lex\_diversity=10 and order\_diversity=10, which control the extent of lexical substitution and sentence-order rearrangement, respectively. HiMark achieves 90.01\% accuracy, compared with 52.38\% for BiMark, 49.79\% for MPAC(1), and 52.92\% for MPAC(1.5), showing that HiMark preserves sufficient bit-level evidence after rewriting.

\subsection{Zero-bit Detection Performance}
Fig.~\ref{fig:zero_bit_detection} shows zero-bit detection performance under different text lengths, measured by true positive rate (TPR) at 1\% false positive rate (FPR). HiMark achieves the highest TPRs across all lengths, indicating reliable zero-bit detection.

\subsection{Ablation Study}
1) Support threshold analysis: Fig.~\ref{fig:ablation_all}(a) shows that increasing $\tau_s$ improves extraction accuracy but increases PPL. A higher threshold removes weak-support candidates and strengthens watermark evidence, but narrows the sampling space. We set $\tau_s=5$ to balance accuracy and PPL.

2) Tournament depth analysis: Fig.~\ref{fig:ablation_all}(b) shows that increasing $m$ from 9 to 12 decreases both extraction accuracy and PPL. A larger $m$ improves fluency by providing more candidates, but weakens watermark selection pressure under fixed refinement and filtering. We set $m=11$ to balance extraction performance and PPL.

3) Refinement size analysis: Fig.~\ref{fig:ablation_all}(c) shows that a smaller $K_{\text{refine}}$ improves extraction accuracy but increases PPL. Fewer candidates strengthen bit-level evidence but reduce fluency. We set $K_{\text{refine}}=10$ to balance extraction strength and text quality.

4) Candidate retention size analysis: Fig.~\ref{fig:ablation_all}(d) shows that a larger $K_{\text{keep}}$ improves extraction by retaining more candidates for refinement, but may increase PPL by selecting tokens less preferred by the LLM. We set $K_{\text{keep}}=32$ to balance robustness and text quality.

\section{Conclusion}
This paper introduces HiMark, a hierarchical and controllable multi-bit watermarking framework for LLM-generated text.  HiMark embeds a binary string through hierarchical candidate control and extracts it in a model-agnostic manner by aggregating multi-layer token evidence in a voting matrix. Experimental results show that HiMark achieves higher watermark extraction accuracy while maintaining text quality and robustness against paraphrasing attacks. 

\section*{Acknowledgment}
This research was supported by the Basic Research Program for Natural Science of Guizhou Province under Grant Number QIANKEHEJICHU-ZD[2025]-043, the Science and Technology Commission of Shanghai Municipality under Grant Number 24ZR1424000, the Nanning ``Yong Jiang'' Program under Grant Number RC20250102, and the National Natural Science Foundation of China under Grant Number U23B2023.

\end{document}